\documentclass[a4paper]{article}
\usepackage{geometry}
\usepackage{graphicx}
\usepackage{graphics}
\usepackage{amsmath}
\usepackage{amssymb}
\usepackage{mathtools}
\usepackage{epstopdf}
\usepackage{eepic}
\usepackage{epsfig}
\usepackage{geometry}
\usepackage{psfrag}
\usepackage[mathscr]{euscript}

\newcommand{\be}{\begin{equation}}
\newcommand{\ee}{\end{equation}}
\newcommand{\ba}{\begin{array}}
\newcommand{\ea}{\end{array}}
\newcommand{\bea}{\begin{eqnarray}}
\newcommand{\eea}{\end{eqnarray}}

\newcommand{\al}{\alpha}
\newcommand{\pa}{\partial}

\newcommand{\la}{\lambda}

\newcommand{\De}{\Delta}
\newcommand{\vphi}{\varphi}

\newcommand{\tha}{\theta}

\newcommand{\rar}{\rightarrow}

\begin{document}

\title{Prolate spheroidal coordinates and the hydrogen molecular ion}

\title{Particle in a field of two centers in prolate spheroidal coordinates:
 integrability and solvability}

\author{Willard Miller, Jr.\\
  School of Mathematics, University of Minnesota, \\
Minneapolis, Minnesota, U.S.A.\\
miller@ima.umn.edu\\[8pt]
and \\[8pt]
Alexander V Turbiner\\
Instituto de Ciencias Nucleares, UNAM, M\'exico DF 04510, Mexico\\
and\\
Institut des Hautes Etudes Scientifique, Bures-sur-Yvette 91440, France\\
turbiner@nucleares.unam.mx, turbiner@ihes.fr}

\maketitle

\begin{abstract}
We analyze one particle, two-center quantum problems which admit separation of variables
in prolate spheroidal coordinates, a natural restriction satisfied by the H$_2^+$ molecular ion. The symmetry operator is constructed explicitly. We give the details of the Hamiltonian reduction of the 3D system to a 2D system with modified potential that is separable in elliptic coordinates. The potentials for which there is double-periodicity of the Schr\"odinger operator in the space of prolate spheroidal coordinates, including one for the  H$_2^+$ molecular ion, are indicated. We study possible potentials that admit exact-solvability is as well as all models known to us  with the (quasi)-exact-solvability property for the separation equations. We find  deep connections between second-order superintegrable and conformally superintegrable systems and these tractable problems. In particular we derive a general 4-parameter expression for a model potential that is always integrable and is conformally superintegrable for some parameter choices.
\end{abstract}

\newpage

\section{Introduction: Symmetry reduction}

Let us consider $3D$ Euclidean space quantum problem in $(x, y, z)$ coordinates with $2D$ potential $V$ which has azimuthal symmetry w.r.t. rotations around $z$-axis. It is convenient to introduce (or parametrize the $3D$ space) the spherical coordinates $(r, \tha, \vphi)$. In these coordinates the Hamiltonian
\begin{equation}
\label{H3r}
   {\cal H}^{(3,r)}\ =\ -\De^{(3)} + V(r, \tha)\ , \ x \in {\bf R}^3 \ ,
\end{equation}
where $\De^{(3)}$ is the $3D$ Laplacian,
\[
  \De^{(3)}= \pa_r^2 + \frac{2}{r}\pa_r + \frac{1}{r^2}\pa_{\tha}^2 +
  \frac{\cot \tha}{r^2}\pa_{\tha}+ \frac{1}{r^2 \sin^2 \tha}\pa_{\vphi}^2.
\]
The problem (\ref{H3r}) admits the symmetry
$L_{\vphi}\ =\ -i\, \pa_{\vphi}$ and dependence on the azimuthal angle ${\vphi}$ can be separated out.
Any eigenfunction has a form $\psi(r, \tha)\ e^{i m \vphi}$, where $m$ is integer (the separation constant or magnetic quantum number).
Separating out the ${\vphi}$-dependence we arrive at the spectral problem for the $2D$ operator,
\[
  {\tilde{\cal H}}^{(3,r)} (r, \tha)\ =\ -\pa_r^2 - \frac{2}{r}\pa_r - \frac{1}{r^2}\pa_{\tha}^2 - \frac{\cot \tha}{r^2}\pa_{\tha} + \frac{m^2}{r^2 \sin^2 \tha}  +
  V(r, \tha)\ ,
\]
with $\psi(r, \tha)$ as eigenfunction. By making the gauge rotation of this operator,
\[
  \rho^{1/2}{\tilde{\cal H}}^{(3,r)}\rho^{-1/2} =  (r \sin \tha)^{1/2} {\tilde{\cal H}}^{(3,r)} (r, \tha) (r \sin \tha)^{-1/2}
\]
one can arrive at the $2D$ Hamiltonian
\begin{equation}
\label{H2r}
   {\cal H}^{(2,r)}\ =\ -\De^{(2)} + \frac{m^2-1/4}{r^2 \sin^2 \tha}  + V(r, \tha),
\end{equation}
where $\De^{(2)}$ is the $2D$ Laplacian,
\[
  \De^{(2)}= \pa_r^2 + \frac{1}{r}\pa_r + \frac{1}{r^2}\pa_{\tha}^2
\]
with domain $r \in [0, \infty)$ and $\tha \in (0, \pi)$, the half-plane. Of course,  similar results would be obtained if we parameterize $3D$ space by ellipsoidal coordinates $(\xi, \eta, \vphi)$
or coordinates $(r_1, r_2, \vphi)$ (see Fig.1). Here the ``interaction" plane $(x, y)$ is parameterized by elliptic coordinates $(\xi, \eta)$ or the coordinates $(r_1, r_2)$, respectively. If prolate spheroidal coordinates are used for $3D$ space parametrization (see below) an important observation should be made: the $3D$ potential depends on $\al, \beta$ only. The gauge rotation is needed to get a $2D$ Hamiltonian. This reduction of a $3D$ spectral problem with azimuthal symmetry to a $2D$ one is a reflection of the representation of $3D$ space as ${\bf R}^3 = {\bf R}^2_+ \times S^1$, where ${\bf R}^2_+$ is half-plane.

Let us consider the $3D$ quantum problem in $(x, y, z)$ coordinates with $2D$ potential $V(x,y)$ which  is translation-invariant w.r.t. $z$-axis, thus, with $3D$ space decomposition ${\bf R}^3 = {\bf R}^2 \times {\bf R}$. In this case the evident symmetry is the momentum $p_z = -i \pa_z$, its eigenfunction is plane wave $\sim e^{i k z}$, thus, the $z$-variable in the $3D$ Hamiltonian can be separated out. For restriction to $2D$ space  a gauge rotation (see above) is unnecessary and the potential in the $2D$ Hamiltonian coincides with the $3D$ potential up to a constant, $k^2$.

Such a procedure of relating $3D$ and $2D$ quantum problems can be thought as analogous to the  the Hamiltonian reduction method in classical mechanics, e.g. \cite{OR}, when the underlying (configuration) space contains a symmetric space as subspace, say,
${\bf R}^n = {\tilde{\bf R}}^m \times Symm^{m-n}$. It is not an instance of the Olshanetsky-Perelomov approach \cite{OP}. See \cite{BCM} and Chapter 4 of \cite{MillerVS} for examples related to separation of variables.

\section{Introduction: two-center problem}\label{section2}

\begin{center}
\begin{figure}
\includegraphics[width=3in,height=2in,angle=0]{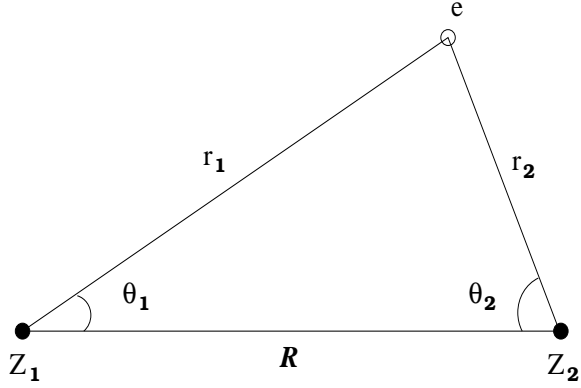}
\caption{\label{Z2+}
Geometric settings for one-particle two-center problem, the reference point is at the middle of the intercenter distance, the particle position at $(x,y,\varphi)$}
\end{figure}
\end{center}

We consider the one-particle, two-center quantum problem characterized by the potential \linebreak
$V(r_1, r_2; R)$ (see Fig.\ref{Z2+}).
There are two possible (related) formulations of the eigenvalue problem: (i) spacial, with the Hamiltonian
\begin{equation}
\label{H3}
   {\cal H}^{(3)}\ =\ -\De^{(3)} + V(r_1, r_2; R)\ , \ x \in {\bf R}^3 \ ,
\end{equation}
where $\De^{(3)}$ is the 3D Laplacian; (ii) planar, with the Hamiltonian
\begin{equation}
\label{H2}
   {\cal H}^{(2)}\ =\ -\De^{(2)} + {\tilde V}(r_1, r_2; R)\ , \ x \in {\bf R}^2 \ ,
\end{equation}
where $\De^{(2)}$ is the 2D Laplacian. The problem (i): $ {\cal H}^{(3)}\Theta=E\Theta$  admits the integral $L_{\vphi}\ =\ -i\, \pa_{\vphi}$ and dependence on the azimuthal angle variable ${\vphi}$ can be separated out. Any eigenstate of ${\cal H}^{(3)}$ is characterized by azimuthal quantum number (magnetic quantum number) $m=0, \pm 1, \pm 2 \ldots$ which is the eigenvalue of $L_{\vphi}$.
The eigenfunctions of ${\cal H}^{(3)}$ have the form
$\Theta=e^{i m \vphi} \rho^{-1/2}\Psi(r_1, r_2)$,
where $\rho^{-1/2}$ is the  gauge factor and  $\Psi(r_1, r_2)$ are the eigenstates of ${\cal H}^{(2)}$ with modified potential
\[
     \tilde V(r_1, r_2; R)\ =\ V(r_1, r_2; R)\ +\ V_{\vphi}(r_1, r_2; R)\ ,
\]
\begin{equation}
\label{VT}
     V_{\vphi} = \frac{(m^2-\frac14)R^2}{r_1 r_2}\bigg[ \frac{1}{(r_1+r_2+R)(r_1+r_2-R)}
     - \frac{1}{(r_1-r_2+R)(r_1-r_2-R)}\bigg]\ .
\end{equation}
Hence, the $3D$ problem (i) with $\Theta\in L^2({\bf R^3})$ is reduced to the $2D$ problem (ii)
with a modified scalar potential $\tilde V(r_1, r_2)$ and $\Psi\in { L}^2({\tilde{\bf R}}^2)$,
where ${\tilde{\bf R}}^2={\bf R}^2_+$ is the upper half plane. In this case
${\bf R}^3 = {\tilde{\bf R}}^2 \times S^1$.
Due to this reduction, it is sufficient to study the planar eigenvalue problem
\[
    {\cal H}^{(2)} \Psi\ =\ E \Psi\ ,\ \Psi \in {\tilde L}^2({\tilde{\bf R}}^2)\ .
\]

 \section{Prolate spheroidal coordinates}

As in \S \ref{section2} we consider  the one-particle, two-center quantum problem - the one-electron diatomic molecular ion $(Z_1,Z_2,e)$ in the Born-Oppenheimer approximation. If $Z_1=Z_2=1$ it becomes the Hydrogen molecular ion $H_2^+$. The Schr\"odinger equation can be written in the Cartesian coordinates as, see e.g. \cite{LL}
\begin{equation}
\label{H2+}
{\cal H}^{(3)}\Theta\equiv (-\De^{(3)} \ -\ \frac{Z_1}{\sqrt{x^2+y^2+(z+a)^2}}-\frac{Z_2}{\sqrt{x^2+y^2+(z-a)^2}})\Theta\ =\ E\Theta,
\end{equation}
where $\De^{(3)}\ =\ \pa_{xx}+\pa_{yy}+\pa_{zz}$ and $Z_{1,2}$ are charges of the fixed centers.
In terms of prolate spheroidal coordinates, see e.g. \cite{MillerVS},
\[ x=a\sinh\alpha \sin\beta \cos\phi\ ,\  y=a\sinh\alpha \sin\beta \sin\phi\ ,\
z=a\cosh\alpha \cos\beta\ ,
\]
where $R=2a$, see Fig.\ref{Z2+}, we have
\be
\label{prolatedelta}
\De^{(3)}\ =\
\ee
\[
\frac{1}{a^2(\cosh^2\al - \cos^2\beta)}\left[(\pa_{\al\al}+\coth\al\pa_\al)\ +\
(\pa_{\beta\beta}+\cot\beta\pa_\beta)\right]
+\frac{1}{a^2\sinh^2\al\sin^2\beta}\pa_{\phi\phi}\ .
\]
The Schr\"odinger equation  now takes the  form
\be
\label{hydrogenion}
(\De^{(3)} \ +\ \frac{(Z_1+Z_2)\cosh\al+(Z_2-Z_1)\cos\beta}{a(\cosh^2\al-\cos^2\beta)})\Theta\ =
\ -E\Theta .
\ee
This follows from the metric expressions
\[
ds^2=dx^2+dy^2+dz^2=a^2(\cosh^2\al-\cos^2\beta)(d\al^2+d\beta^2)+a^2\sinh^2\alpha\sin^2\beta\ d\phi^2\ .
\]
For future use we note that $\cosh^2\al - \cos^2\beta=\sinh^2\al+\sin^2\beta$.
Clearly variables separate in equation (\ref{hydrogenion}).

Now we investigate the most general potential $V(\alpha,\beta)$ such that the equation
\be
\label{genform}
 {\cal H}^{(3)} \Theta \equiv (-\De^{(3)} + V(\al,\beta))\Theta = E\Theta
\ee
is separable in  prolate spheroidal coordinates and determine the symmetry operator that implements this variable separation. In order to find it, let we multiply the Hamiltonian ${\cal H}^{(3)}$ in (\ref{genform}) by $(\cosh^2\al-\cos^2\beta)$. We arrive at a generalized spectral problem for the operator
\[
   {\tilde {\cal H}}\ \equiv \ (\cosh^2\al-\cos^2\beta){\cal H}^{(3)}\ =\
\]
\be
\label{Hgen}
 -\frac{1}{a^2}\left[(\pa_{\al\al}+\coth\al\pa_\al)\ +\ (\pa_{\beta\beta}+\cot\beta\pa_\beta) +
  \big(\frac{1}{\sinh^2\al}+ \frac{1}{\sin^2\beta}\big)\pa_{\phi\phi}\right] \ +\
  (\cosh^2\al-\cos^2\beta)V(\al,\beta) \ ,
\ee
with the weight $(\cosh^2\al-\cos^2\beta)$. From expression (\ref{Hgen}) it follows immediately that
\[
V(\al,\beta)=\frac{f(\al)+g(\beta)}{a(\cosh^2\al-\cos^2\beta)}\ ,
\]
so
\be
\label{sepham}
 \bigg(-\De^{(3)} + \frac{f(\al)+g(\beta)}{a(\cosh^2\al - \cos^2\beta)}\bigg)\,\Theta \ = \ E\, \Theta \ ,
\ee
where $f(\al), g(\beta)$ are arbitrary functions.
For the $(Z_1, Z_2, e)$ system those functions are
\be
\label{ZZe}
   f(\al)= {(Z_1+Z_2)} \cosh\al\ ,\ g(\beta) = {(Z_2-Z_1)}\cos\beta\ .
\ee
This implies that the Hamiltonian of the $(Z_1, Z_2, e)$ system in $\al, \beta$ variables is double-periodic, i.e., invariant under the transformation: $\al \rar \al + 2i\pi, \beta \rar \beta + 2 \pi$.
It is worth emphasizing
if, in general,
\be
\label{Vperiod}
    f(\al)\ =\ {\cal F}(\cosh\al)\ ,\ g(\beta)\ =\ {\cal G}(\cos\beta)\ ,
\ee
the Hamiltonian
\[
    {\cal H}^{(3)}\ =\ -\De^{(3)} + \frac{{\cal F}(\cosh\al) + {\cal G}(\cos\beta)}
    {a(\cosh^2\al - \cos^2\beta)}\ ,
\]
is double-periodic: $\al \rar \al + 2i\pi, \beta \rar \beta + 2 \pi$.

Writing $\Theta$ in separable form $\Theta=A(\al)B(\beta) \varPhi(\phi)$, we see that the spectral problem for the operator (\ref{Hgen}) (see also (\ref{genform})) separates as
\be
\label{sepeqn1}
A''(\al)+\coth\al A'(\al)-a f(\al)A(\al)+
\left[-\frac{m^2}{\sinh^2\al}-a^2\la +a^2 E \sinh^2\al \right]A(\al)\ =\ 0\ ,
\ee
\be
\label{sepeqn2}
B''(\beta)+\cot\beta B'(\beta)
- a g(\beta)B(\beta)+\left[-\frac{m^2}{\sin^2\beta} + a^2 \la
+ a^2 E \sin^2\beta \right]B(\beta) = 0\ ,
\ee
\be
\label{sepeqn3}
C''(\phi) + m^2C(\phi) = 0\ .
\ee
Here, $m,\la,E$ are the separation constants. It is worth noting that the separation equations similar to (\ref{sepeqn1}), (\ref{sepeqn2}) appear when the separation of variables for Riemannian spaces of constant curvature (with 0 potential) is studied \cite{Kalnins}, \cite{KKMP}. With $m$ fixed, the equations (\ref{sepeqn1}), (\ref{sepeqn2}) define a bi-spectral problem with spectral parameters $\la,E$. Making a gauge transformation we can remove the second term in (\ref{sepeqn1}), (\ref{sepeqn2})) and reduce, hence, (\ref{sepeqn1}), (\ref{sepeqn2}))  to one-variable  Schr\"odinger equation form. Solving for $\la$ in these equations we obtain
\[
\la \Theta=\frac12\left[\frac{1}{a^2}(\pa_{\al\al}+\coth\al\pa_\al)\ -\ \frac{f(\al)}{a}
-\frac{1}{a^2}(\pa_{\beta\beta}+\cot\beta\pa_\beta)
\ +\ \frac{g(\beta)}{a}
\right.
\]
\[
\left.
+ \frac{1}{a^2}(\frac{1}{\sinh^2\al}- \frac{1}{\sin^2\beta})\pa_{\phi\phi}+(\sinh^2\al\ -\ \sin^2\beta)E\right]\Theta\ ,
\]
where $\Theta=A B \varPhi$. Using equations (\ref{sepham}),(\ref{prolatedelta}) to solve for $E\,\Theta$ and substitute into this expression, we find
\[
K\,\Theta = \la\,\Theta\ ,
\]
where the operator
\be\label{Koperator}
 K\ =\ \frac{1}{a^2(\sinh^2\al+\sin^2\beta)}\
   \bigg(\sin^2\beta\ (\pa_{\al\al}\ +\ \coth\al\pa_\al)\ -\
   \sinh^2\al\ (\pa_{\beta\beta}+\cot\beta\pa_\beta)\bigg)
\ee
\[
   + \frac{\sin^2\beta - \sinh^2\al}{a^2\sinh^2\al\sin^2\beta}\ \pa_{\phi\phi}\ -\
   \frac{\sin^2\beta\ f(\al)\ -\ \sinh^2\al\ g(\beta)}{a(\sinh^2\al\ +\ \sin^2\beta)}\ .
\]
In the case of the $(Z_1, Z_2, e)$ system this operator coincides with the integral found by Erikson and Hill \cite{Erikson-Hill:1949}.

It follows from the general theory of variable separation \cite{MillerVS} that
$
      [K,H]=0$,
so that $K$ is a symmetry operator for the system. Moreover, the pure differential operator part of $K$ can be expressed in terms of the
enveloping algebra of the Euclidean Lie algebra $e(3)$. A basis for $e(3)$ is given by the 3 translation generators
\[
     P_1=\pa_x,\ P_2=\pa_y,\ P_3=\pa_z,
\]
and the 3 rotation generators
\[
     J_1=y\pa_z-z\pa_y,\ J_2=z\pa_x-x\pa_z,\ J_3=x\pa_y-y\pa_x,
\]
in Cartesian coordinates. In terms of these generators we find
\[
-{\cal H}\ =\ P_1^2+P_2^2+P_3^2 - \frac{f(\al)+g(\beta)}{a(\cosh^2\al-\cos^2\beta)}\ ,
\]
\[
K=-\frac{1}{a^2}(J_1^2+J_2^2+J_3^2-a^2(P_1^2+P_2^2))
-\frac{f(\al)+g(\beta)}{a(\cosh^2\al-\cos^2\beta)}
+\frac{\cos^2\beta\ f(\al)+\cosh^2\al\ g(\beta)}{a(\cosh^2\al-\cos^2\beta)}\ .
\]

So far we studied the spacial problem (i) characterized by the Hamiltonian (\ref{H3}).
Now we can look what would happen in planar motion formalism (ii), where the problem is described by the Hamiltonian (\ref{H2}),
when we separate out azimuthal motion, coordinate $\vphi$, and perform the gauge transformation $\rho=(\sinh\alpha\ \sin\beta)^{-1/2}$.
The $2D$ Laplacian becomes
\[
\frac{1}{a^2(\cosh^2\al - \cos^2\beta)}\left[\pa_{\al\al} +\
\pa_{\beta\beta}\right] \ =\ \De^{(2)}\ .
\]
For the case of the $(Z_1,Z_2,e)$ system the Schr\"odinger equation is
\be
\label{hydrogenion2}
\bigg\{\De^{(2)} \ +\ \frac{\bigg[\frac{\frac14-m^2}{\cosh^2\alpha-1}+(Z_1+Z_2)\cosh\al\bigg]+
 \bigg[\frac{\frac14-m^2}{1-\cos^2\beta}+(Z_2-Z_1)\cos\beta\bigg]}{a(\cosh^2\al-\cos^2\beta)}\bigg\}
 \Psi\ =\ -E\Psi\ .
\ee
(cf. (\ref{H2})). The metric length is
\[
ds^2=dx^2+dy^2=a^2(\cosh^2\al-\cos^2\beta)(d\al^2+d\beta^2)\ .
\]
The most general $2D$ Schr\"odinger equation which admits separation of variables in $\al, \beta$ coordinates has the form
\be
\label{sepham2}
 \bigg(-\De^{(2)} + \frac{f(\al)+g(\beta)}{a(\cosh^2\al-\cos^2\beta)}\bigg)\Psi \ = \ E\Psi\ .
\ee
where $f(\al), g(\beta)$ are arbitrary functions.
Assuming $\Psi=A(\al)B(\beta)$ we arrive at two equations
\be
\label{sepeqn1-2}
A''(\al)\  -\ a f(\al)A(\al)\ +\left[- a^2\la\
+\ a^2 E \sinh^2\al\ \right]A(\al)\ =\ 0\ ,
\ee
\be
\label{sepeqn2-2}
B''(\beta) \  -\ a g(\beta)B(\beta)\ + \left[a^2\la
\ +\ a^2 E \sin^2\beta\ \right]B(\beta)\ =\ 0\ ,
\ee
where $\la$ is the separation parameter. These two equations represent a bi-spectral problem where $E, \la$ are spectral parameters.

The operator
\be\label{Koperator2}
K^{(2)}\ =\ \frac{1}{a^2(\sinh^2\al+\sin^2\beta)}\bigg(\sin^2\beta\ \pa_{\al\al}
    \ -\ \sinh^2\al\  \pa_{\beta\beta}\bigg)
    \ -\ \frac{\sin^2\beta\ f(\al)-\sinh^2\al\ g(\beta)}{a(\sinh^2\al+\sin^2\beta)}\ ,
\ee
(cf. (\ref{Koperator})), commutes with the planar Hamiltonian ${\cal H}^{(2)}$,
\[
[K^{(2)} , {\cal H}^{(2)}]\ =\ 0\ .
\]
In the case of the $(Z_1, Z_2, e)$ system the operator $K^{(2)}$ coincides with the integral found by Erikson and Hill \cite{Erikson-Hill:1949}.

\section{Elliptic coordinates}
In terms of the more physical (dimensionless) elliptic coordinates, see e.g. \cite{LL},
\be
\label{xieta}
\xi=\cosh\al\ ,\  \eta=\cos\beta\ ,
\ee
where $\xi \in [1, \infty)\ ,\ \eta \in [-1,1]$,
which implies invariance with respect to translations $\al \rar \al + 2i\pi,
\beta \rar \beta + 2 \pi$, respectively,
the planar Hamiltonian for the system $(Z_1, Z_2, e)$ takes the form
\[
{\cal H}^{(2)} \Psi \equiv (-\De^{(2)} - \frac{Z_1}{r_1} - \frac{Z_2}{r_2}+V_\varphi)\Psi\ =\ E \Psi\ ,
\]
(see (\ref{VT})), where $r_1=a(\xi+\eta)$, $r_2=a(\xi-\eta)$, see Fig.1, and
\be
\label{prolatedeltaxi}
  \De^{(2)}\ =\ \frac{\xi^2-1}{a^2(\xi^2-\eta^2)}[\pa_{\xi\xi}+\frac{\xi}{\xi^2-1}\pa_\xi] +
  \frac{1-\eta^2}{a^2(\xi^2-\eta^2)} [\pa_{\eta\eta}-\frac{\eta}{1-\eta^2}\pa_\eta]\ .
\ee

The general separable form for the Hamiltonian eigenvalue equation in $\xi, \eta$ coordinates is
\be
\label{Hxieta}
 {\cal H}^{(2)} \Psi \equiv \bigg(-\De^{(2)} + \frac{{\cal F}(\xi) + {\cal G} (\eta)} {a^2(\xi^2-\eta^2)}\bigg)\Psi\ =\ E \Psi\ ,
\ee
(cf. (\ref{Vperiod})). This form naturally includes the modified potential $V_{\vphi}(r_1, r_2)$ (see (\ref{VT})), which occurs in a transition from $3D$ case (ii) to $2D$ case (i), see Section 1,
\[
     V_{\vphi}(\xi, \eta) \ =\ \frac{m^2-\frac14}{a^2}\bigg(\frac{1}{\xi^2-1}+\frac{1}{1-\eta^2}\bigg)
     \frac{1}{\xi^2-\eta^2}\ ,
\]
(cf. (\ref{VT})).
We emphasize that in the case of the $(Z_1, Z_2, e)$ system the functions in the Hamiltonian (\ref{Hxieta}) take the  simple form
\[
   {\cal F}(\xi) = {(Z_1+Z_2)} \xi\ +\frac{m^2-\frac14}{a (\xi^2-1)}\quad ,\quad {\cal G} (\eta) = {(Z_2-Z_1)}\eta\ +\frac{m^2-\frac14}{a (1-\eta^2)}\ .
\]
In general, the one-particle, two-center potential $V(r_1, r_2)$ in (\ref{H3}), which admits separation of variables in ellipsoidal coordinates, has the form
\be
\label{Vr1r2}
    V(r_1, r_2)\ =\ \frac{f(r_1 + r_2) - g(r_1 - r_2)}{r_1}\ +\
    \frac{f(r_1 + r_2) + g(r_1 - r_2)}{r_2} \ ,
\ee
where $f, g$ are arbitrary functions. In the case of $(Z_1, Z_2, e)$ system $f, g$ are constants. Interestingly, if $f, g$ are linear functions, $f(x) = A x, g(x) = B x$, the potential (\ref{Vr1r2}) becomes,
\[
     V(r_1, r_2)\ =\ (A+B)\bigg(\frac{r_1}{r_2} + \frac{r_2}{r_1} \bigg)\ .
\]
Multiplying (\ref{Hxieta}) by $(\xi^2-\eta^2)$ and writing $\Psi$ in separable form $\Psi={\cal A}(\xi){\cal B}(\eta)$, we obtain the separation equations
\be
\label{sepeqn1a}
(\xi^2-1) {\cal A}'' + \xi{\cal A}' + a{\cal F}(\xi){\cal A} + \left[-a^2\la + a^2\xi^2E\right]{\cal A}\ =\ 0\ ,
\ee
\be
\label{sepeqn2a}
(1-\eta^2){\cal B}'' - \eta{\cal B}' + a{\cal G}(\eta){\cal B} + \left[ a^2\la - a^2\eta^2 E\right]{\cal B}\ =\ 0\ ,
\ee
(cf. (\ref{sepeqn1}), (\ref{sepeqn2})), where  $\la, E$ are the separation constants playing a role of spectral parameters in bi-spectral problem (\ref{sepeqn1a})-(\ref{sepeqn2a}).
Now the integral takes the form
\be
\label{Koperatora}
K=\frac{1}{a^2(\xi^2-\eta^2)}\bigg((\xi^2-1)(1-\eta^2)\pa_{\xi\xi}
-\xi(\eta^2-1)\pa_\xi+(\xi^2-1)(\eta^2-1)\pa_{\eta\eta}+\eta(\xi^2-1)\pa_\eta\bigg)
\ee
\[
+\frac{(1-\eta^2){\cal F}(\xi)+(1-\xi^2){\cal G}(\eta)}{a(\xi^2-\eta^2)}\ ,
\]
(cf. (\ref{Koperator})) and the separation constant $\la$ is the spectral parameter in the eigenvalue problem $K \Psi = \la \Psi$.

\section{Solvability}

The goal of the section is to describe one-particle, two-center potentials for which  exact solutions can be found. In order to proceed
let us take the equation (\ref{sepeqn1a}) in the form of an eigenvalue problem
\be
\label{sepeqn1b}
h_{\xi} {\cal A} \equiv \bigg( -(\xi^2-1) \pa^2_{\xi} - \xi \pa_{\xi} - a {\cal F}(\xi) - a^2E \xi^2\bigg) {\cal A}\ =\ -a^2\la {\cal A}\ .
\ee
Written in terms of the variable $\al$ where $\xi=\cosh\al$, this becomes a 1D Schr\"odinger equation for the Hamiltonian
\[
   {\cal H}_{\xi}(\al)\ =\ -\pa^2_{\al} + V_{\xi}(\xi=\cosh \al)\ ,
\]
with potential
\be
\label{Vxi}
   V_{\xi}\ =\ - a{\cal F}(\xi) - a^2E \xi^2=\ -a{\cal F}(\cosh\al)-a^2E\cosh^2\al\ ,
\ee
and we arrive at the eigenvalue problem for the Schr\"odinger operator with a hyperbolic potential. (In most of the following discussion we restrict our attention to a single energy eigenspace so we can rescale matters so that $E=0$.) Among hyperbolic potentials there is the exactly-solvable hyperbolic modified P\"oschl-Teller potential (in other words, one-soliton potential for $A_s=0$)
\be
\label{VhPT}
 V^{(h)}_{PT}\ =\ -\frac{A_c}{\cosh^2 \al}-\frac{A_s}{\sinh^2 \al}\ ,
\ee
with a finite number of bound states, where all of them can be found exactly(algebraically), see e.g. \cite{LL},
and two three-parametric families \footnote{The fourth parameter should take discrete values} of quasi-exactly-solvable potentials:
\be
\label{Vh1}
   V^{(h, qes)}_1\ =\ -\frac{A_c}{\cosh^2 \al} -\frac{A_s}{\sinh^2 \al} + A_1{\cosh^2 \al} + A_2{\cosh^4 \al}
   \ ,
\ee
and
\be
\label{Vh2}
   V^{(h, qes)}_2\ =\ -\frac{A_c}{\cosh^2 \al} -\frac{A_s}{\sinh^2 \al} + \frac{A_1}{\cosh^4 \al} + \frac{A_2}{\cosh^6 \al}
   \ ,
\ee
where a finite number of eigenstates can be found algebraically \cite{Turbiner:1988}. (These potentials are all related to superintegrable systems, see, e.g., \cite{KMP2}, eqn. (72).)  
Known eigenfunctions for (\ref{VhPT}) have a form
\[
\cosh^{\nu}(\frac{\al}{2})\ \sinh^{\mu}(\frac{\al}{2})\ P_k (\cosh \al)\ ,
\]
where $\nu, \mu$ are known constants defined by $A_{c,s}$ and $P_k$ is a polynomial of degree $k$, which can be found by algebraic means. Known (algebraic) eigenfunctions for (\ref{Vh1}) have a form
\[
\cosh^{\nu}(\frac{\al}{2})\ \sinh^{\mu}(\frac{\al}{2}) \ e^{-b \cosh \al}\ P_k (\cosh \al)\ ,
\]
where $b^2=A_2$. Similar form (up to a factor) appears for the potential (\ref{Vh2}).

The same analysis as for (\ref{sepeqn1a}) can be performed for the equation (\ref{sepeqn2a}) rewriting it like
\[
h_{\eta} {\cal B} \equiv \bigg( -(1-\eta^2) \pa^2_{\eta} + \eta \pa_{\eta} - a {\cal G}(\eta) + a^2E \eta^2\bigg) {\cal B}\ =\ a^2\la {\cal B}\ .
\]
Indeed,  we arrive at the eigenvalue problem for the Hamiltonian
\[
   {\cal H}_{\eta}(\beta)\ =\ -\pa^2_{\beta} + V_{\eta}(\eta=\cos \beta)\ ,
\]
with trigonometric potential,
\be
\label{Veta}
   V_{\eta}\ =\ - a{\cal G}(\eta) + a^2E \eta^2=\ -a{\cal G}(\cos\beta) + a^2E\cos^2\beta\ .
\ee
Among trigonometric potentials there is the exactly-solvable modified trigonometric P\"oschl-Teller potential (see e.g. \cite{Turbiner:2013})
\be
\label{VtPT}
    V^{(t)}_{PT}\ =\ -\frac{B_c}{\cos^2 \beta} - \frac{B_s}{\sin^2 \beta}\ ,
\ee
and also two three-parametric families \footnote{The fourth parameter should take discrete values} of quasi-exactly-solvable potentials
\be
\label{Vt1}
   V^{(t,qes)}_1\ =\ -\frac{B_c}{\cos^2 \beta} - \frac{B_s}{\sin^2 \beta} + B_1{\cos^2 \beta} + B_2{\cos^4 \beta}\ ,
\ee
and
\be
\label{Vt2}
   V^{(t,qes)}_2\ =\ -\frac{B_c}{\cos^2 \beta} - \frac{B_s}{\sin^2 \beta} + \frac{B_2}{\cos^4 \beta} + \frac{B_3}{\cos^6 \beta}\ ,
\ee
where a finite number of eigenstates can be found algebraically. Known eigenfunctions for (\ref{VtPT}) have a form
\[
|\cos(\frac{\beta}{2})|^{\nu}\ |\sin(\frac{\beta}{2})|^{\mu} P_k (\cos \beta)
\]
where $\nu, \mu$ are known constants defined by $B_{c,s}$ and $P_k$ is a polynomial of degree $k$, which can be found by algebraic means. Known (algebraic) eigenfunctions for (\ref{Vt1}) have a form
\[
|\cos(\frac{\beta}{2})|^{\nu}\ |\sin(\frac{\beta}{2})|^{\mu}\ e^{-b \cos \beta}\ P_k (\cos \beta)
\]
where $b^2=B_2$. Similar form (up to a factor) appears for the potential (\ref{Vt2}).

Finding ${\cal F}, {\cal G}$ from (\ref{VhPT}), (\ref{VtPT}) we construct the  two-dimensional, four-parametric, exactly solvable problem with potential
\be
\label{exactly}
     V_{PT}(\xi, \eta)\ =\ \frac{1}{\xi^2 - \eta^2} \bigg(\frac{A_c}{\xi^2}+\frac{A_s}{1 - \xi^2} +
     \frac{B_c}{\eta^2} - \frac{B_s}{1-\eta^2}\bigg)\ ,
\ee
see (\ref{Hxieta}). If $B_c \neq 0$ the potential is singular. In $(r_1,r_2)$ variables the potential (\ref{exactly}) looks as follows
\be
\label{exactly12}
     V_{PT}(r_1,r_2)\ =\ \frac{1}{r_1 r_2} \bigg(\frac{a_c}{(r_1+r_2)^2}+\frac{a_s}{(r_1+r_2)^2-R^2} +
     \frac{b_c}{(r_1-r_2)^2} - \frac{b_s}{(r_1-r_2)^2-R^2}\bigg)\ .
\ee
This model is integrable with one  second-order integral (\ref{Koperatora}) (with appropriate ${\cal F}, {\cal G}$) for any values of parameters $a_{c,s}, b_{c,s}$. As we will point out, for 2 linear conditions on the 4 parameters it is conformally second-order superintegrable. The question of the existence of the second, higher-order-than-two integral (thus superintegrability) for certain values of parameters is open and might be a subject of separate investigation.

This construction is similar to one which has led to TTW model (see \cite{TTW:2009}), when the separation in polar coordinates was inverted by adding 2D radial harmonic oscillator to modified trigonometric P\"oschl-Teller potential in angular coordinate (\ref{VtPT}).
Taking in (\ref{Hxieta}) a superposition of (quasi)-exactly-solvable, hyperbolic-trigonometric potentials we will obtain (quasi)-exactly-solvable, integrable, one-particle, two-center problems.

There is single 2D Euclidean space nondegenerate superintegrable system that permits separation in elliptic coordinates, the Smorodinsky-Winternitz system (or caged oscillator) \cite{Fris:1965, Wint:1966}:
\be
\label{SW}
-\De^{(2)}\Psi + \left(A_1(x^2+y^2) + \frac{A_2}{x^2} + \frac{A_3}{y^2}\right)\Psi\ =\ E\Psi\ .
\ee
or, in other words, TTW model at $k=1$ \cite{TTW:2009}. In this case
\be
{\cal F}(\xi)=a^4A_1\xi^2(\xi^2-1)+\frac{A_2}{\xi^2-1}-\frac{A_3}{\xi^2},\quad
{\cal G}(\eta)=-a^4A_1\eta^2(\eta^2-1)+\frac{A_2}{1-\eta^2}+\frac{A_3}{\eta^2}\ .
\ee
It has two second order integrals and the expressions are valid for all $E$ simultaneously. 
If we restrict $E$ to a fixed value, say $E=0$, then we can consider conformal symmetries of 
the Schr\"odinger operator.  The system (\ref{SW}) becomes now trivially second-order conformally superintegrable with 3 generators. Note that  for $A_1=0$ the model (\ref{SW}) degenerates to (\ref{exactly}). However, this is not {\it just}  a restriction because the restricted system 
is now conformally second-order  superintegrable with 6 linearly independent generators, see \cite{KKMP2}, eqn. (2), so the symmetry algebra is much larger.
Hence, the model with (\ref{exactly}) for a certain particular values of of parameters is conformally  superintegrable. Another way that (\ref{SW}) leads to (\ref{exactly}), but with different values of the parameters, is that it is conformally equivalent to a second-order superintegrable system on the 2-sheet hyperboloid, see \cite{KMP}, eqns.\ (4)-(9). Again the 1D potentials are of  P\"oschl-Teller type.

\section*{Acknowledgments}

  A.V.T. is thankful to University of Minnesota for kind hospitality extended to him where this work was initiated.
  The first author was partially supported by a grant from the Simons Foundation (\# 208754 to Willard Miller, Jr.).
  The second author is supported in part by the University Program FENOMEC, and by the PAPIIT
  grant {\bf IN109512} and CONACyT grant {\bf 166189}~(Mexico).

\end{document}